\documentclass[12pt,preprint]{aastex}
\usepackage{epsf}
\usepackage{lscape}
\usepackage{graphicx}
\usepackage{natbib}
\begin{document}
\title{Schwarzschild Lecture 2014: HectoMAPping the Universe }
\author{Margaret J. Geller$^1$ \and Ho Seong Hwang$^2$}
\affil{$^1$Harvard-Smithsonian Center for Astrophysics\\
$^2$Korean Institute for Advanced Study}

\begin{abstract}During the last three decades progress in mapping the universe from an age of 400,000 years to the present has been stunning. Instrument/telescope combinations have naturally determined the sampling of various redshift ranges. Here we outline the impact of the Hectospec on the MMT on exploration of the universe in the
redshift range 0.2 $\lesssim z \lesssim 0.8$. We focus on dense redshift surveys, SHELS and HectoMAP. SHELS is a complete magnitude limited survey covering 8 square degrees. The HectoMAP survey  combines a red-selected dense redshift survey and a weak lensing map covering 50 square degrees. Combining the dense redshift survey with a Subaru HyperSuprimeCam (HSC) weak lensing map will provide a powerful probe of the way galaxies trace the distribution of dark matter on a wide range of physical scales. 
\end{abstract}

\section {Introduction}
Names often have deep meaning in our lives and the name Schwarzschild has a special significance in my scientific career. When I entered graduate school in the Princeton physics department in 1970, the environment was difficult for everyone, but it was particularly difficult for women. Princeton had only begun admitting women to the undergraduate school in 1969, and a woman had never received a PhD in physics from Princeton. Fortunately there were a few faculty at Princeton who were kind and encouraging. Martin Schwarzschild was one of those. He was a scientist's scientist who elegantly defined a field. He was also an
insightful, gentle person who, with a few well-chosen words, could ease the feeling of not belonging.

I first met Martin soon after my arrival at Princeton. One day he came to a physics colloquium. I proudly introduced
Martin to one of my contemporaries. My friend said,``Oh, I am so glad to meet you. I know all about your work on black holes.'' With a characteristic twinkle in his eye and in the high-pitched voice I had come to know well,
Martin said ``I'm not that old. That was my father.'' Martin was 58 at that time, younger than I am now. 

Then as now, astrophysics at Princeton was the domain of both the physics department and the department of astrophysical sciences (where Martin Schwarzschild was a professor). By the end of my first year of graduate school I had begun to do research with Jim Peebles. He was laying out the foundations of physical cosmology. I was fascinated by the big questions and I was amazed at how little was known about galaxies and the universe. Peebles started me on the track of using the distribution of galaxies to learn about the parameters that define the evolution of the universe and to explore the way that structure in the universe forms and evolves.

In the early 1970's galaxy redshifts were a precious commodity; very few had been measured.
A catalog of 527 redshifts (large for the time) was the basis for the second paper I wrote with Peebles. We used the redshifts along with Peebles' 1000-particle (!) n-body simulation to make the first statistical virial theorem estimate of the masses of galaxies (Geller \& Peebles 1972). Marc Davis, John Huchra, and I later used a somewhat larger and more complete catalog to estimate the mean mass density of the universe (Davis, Geller \& Huchra 1978). We arrived at
0.2 $\leq \Omega_0 \leq$ 0.7. The range is broad, but even with the small survey of the time, it seems remarkable that the current best value is within our range (Planck Collaboration et al. 2014).

Now it is remarkably easy to measure redshifts for several thousand galaxies in a night. Typical observations reach much deeper than the ones we analyzed in the 1970s. I have been fortunate that my scientific career coincided with the time when we could first map the universe
from the epoch of recombination to the present day. 

Rather than review my role in these remarkable developments here, I prefer to concentrate on recent projects and to preview a large survey, HectoMAP, that I hope to complete in the next few years. Ho Seong Hwang has worked with me on these projects. The introduction and conclusions of this paper are my voice alone: the rest of this paper is a collaboration with Ho Seong Hwang.
 
Over the years following the pioneering CfA redshift surveys (Davis et al. 1982; Geller \& Huchra 1989), telescope and instrument combinations have, at least in part, defined the design and extent of the fantastic array of projects that now define what we know about the structure of our universe. Here we review some of the surveys enabled by the Hectospec on the MMT (Section 2). We use SHELS, a complete magnitude limited survey, to demonstrate
some of the impact of spectroscopic and/or color selection  (Section 3).
We then preview the red-selected HectoMAP project, an international collaboration that combines a large Hectospec redshift survey with a weak lensing map based on Subaru HSC imaging
(Section 4).  The emphasis on combining a dense redshift survey with a weak lensing map extends our pilot surveys of individual clusters (Geller et al. 2014a; Hwang et al. 2014) and of sizable weak lensing fields (Geller et al. 2005; Kurtz et al. 2012). Details of the
HectoMAP results described here will be published in full elsewhere. We conclude in Section 5. We adopt H$_0$ = 70 km s$^{-1}$ Mpc$^{-1}$, $\Omega_\Lambda$ = 0.7, and $\Omega_m$ = 0.3 throughout. 

\section{The Hectospec Redshift Surveys}

During the last 20 years, wide-field multi-object spectrographs on 2.5 to 10-meter telescopes
have enabled a suite of redshift surveys covering a large fraction of the nearby universe(e.g. Shectman et al. 1996; Colless et al.
2001; Jones et al. 2009: Baldry et al. 2010; Ahn et al. 2014) and probing the evolution of the galaxy distribution to redshifts $\gtrsim 1$. Le F´evre et al. (2013) summarize (their Figures 24 through 26 and associated references) the characteristics of the existing surveys along with some future projects.

Here we concentrate on the moderate depth surveys we have carried out with the Hectospec
on the MMT. The Hectospec is a fiber instrument with a 1$^\circ$ degree field of view (Fabricant et al. 2005). There are 300 fibers deployable over the field. For general applications to redshift surveys of clusters of galaxies or of the general galaxy distribution, 250 of the fibers can be placed on survey objects; most of the remaining fibers are allocated to the sky. In the ``sweet'' spot for operation of the instrument typical exposures are 0.75-1.5 hours. The redshift yield can easily be 2000 galaxies per night for galaxies  at a median redshift of $ z = 0.3$. 

The Hectospec aperture is well-matched to surveys of rich clusters of galaxies in the redshift
range 0.1 $\lesssim z \lesssim 0.3$. For example, HeCS (Hectospec Cluster Survey;
Rines et al. 2013) is a
$\sim 20,000$ galaxy survey of 58 clusters in this redshift range. The survey provides measurements of the mass profiles of these clusters over a large radial range that includes the infall region. The infall region lies between the radius where the cluster is virialized and the turnaround radius where the infall velocity induced by the cluster mass concentration just compensates the Hubble flow. In a $\Lambda$CDM universe, the mass currently within the infall region is a good estimate of the mass that the cluster will have in the far future, the ultimate mass of the cluster (Nagamine \& Loeb 2003; Busha et al. 2005; D\"unner et al. 2006). Rines et al. (2013) demonstrate that the observed ultimate  masses are in excellent agreement with current models. This comparison is a new observational test of the
$\Lambda$CDM paradigm.

The densest cluster redshift surveys carried out with Hectospec are particularly powerful when combined with weak lensing maps. As a pilot project, Hwang et al. (2014) cross-correlate
dense redshift surveys for nine well-sampled clusters with weak lensing maps derived from Subaru data (Okabe et al. 2010). Comparison of the maps reveals the impact of superimposed structures
along the line-of-sight on the lensing map. The most serious contributors to the excess lensing signal tend to be structures near the cluster redshift. These superimposed structures contribute an excess lensing signal. When the fractional excess in the cross-correlation at zero lag exceeds $\sim 10$\%, the constraints on the cluster mass tend to be poor; thus the excess cross-correlation signal is an interesting test of the reliability of the weak lensing mass.

Combining a weak lensing map with a foreground redshift survey has been an important motivation for  Hectospec redshift surveys (Geller et al. 2005; Geller et al. 2010; Kurtz et al. 2012). The weak lensing maps of the Deep Lens Survey (DLS) originally motivated the foreground SHELS (Smithsonian Hectospec Lensing Survey) redshift survey of the DLS F2 field. Tests of the projected matter distribution revealed by weak lensing against the distribution derived from the redshift survey show that the galaxies trace the three-dimensional large-scale matter distribution throughout the redshift range of the survey
(Geller et al. 2005). Later using both the DLS map and a field observed in better seeing with the Subaru telescope, Geller et al. (2010), Kurtz et al. (2012), and Utsumi et al. (2014) demonstrate that thresholds for reliable detection of massive clusters of galaxies based on lensing maps have been overly optimistic. These results are consistent with the conclusions of larger surveys
(e.g. CFHTLenS; Shan et al. 2012; 2014). Here we use the complete SHELS survey to set the stage for the somewhat deeper, substantially more extensive HectoMAP survey.

The HectoMAP survey once again combines a dense redshift survey with weak lensing observations. Geller, Diaferio \& Kurtz (2011) preview the 50 square degree survey and outline the initial goals. Here we give more details of the galaxy selection for the redshift survey compared with the SHELS complete surveys. We also preview the first comparisons between the HectoMAP data and the results of simulations (Section \ref{simulation}).

\section{SHELS: A Dense, Complete Magnitude Limited Survey}
 
SHELS is a dense redshift survey covering 8 square degrees of the Deep Lens Survey(DLS: Wittman et al. 2006). Currently SHELS is the densest survey to its limiting apparent magnitude, R = 20.6. The two
widely separated 4 square degree fields of SHELS (the F1 and F2 fields of the DLS) are remarkably different in average galaxy density to the limiting apparent magnitude: the F2 field contains 13408 
extended objects above the magnitude limit whereas the F1 field contains only $\sim$9800
extended objects. The redshift survey of the F1 field is not yet complete, but a comparison 
of the F1 and F2 fields will be an interesting probe of cosmic variance.
Here we focus on the complete F2 field (Geller et al. 2014b).

The 95\% completeness of the F2 redshift survey to its limiting R = 20.6 is greater than the typical completeness of the SDSS Main Galaxy Sample (Strauss et al. 2002; Park \& Hwang 2009). As the SDSS has amply demonstrated, sufficiently large, complete redshift surveys have a wide variety of applications for exploring the 
properties of galaxies and large-scale structure. So far, we have used the F2 SHELS 
survey to explore the H$\alpha$ luminosity function (Westra et al. 2010), the properties of 
star-forming galaxies detected by the WISE satellite at 22$\mu$m (Hwang et al. 2012), and the mass-metallicity relation for galaxies at 0.2$\lesssim z \lesssim0.38$ (Zahid et al. 2013; Zahid et al. 2014).

The F2 survey provides a benchmark for demonstrating the qualitative effects of color selection on
the observable properties of large-scale structure at intermediate redshift. We also comment briefly on the importance of spectroscopic (as opposed to photometric) redshifts as a route to understanding the galaxy distribution and the relationship between the properties of galaxies and their surroundings.

\begin{figure}
\centering
\begin{tabular}{cc}
\includegraphics[width=3.15in]{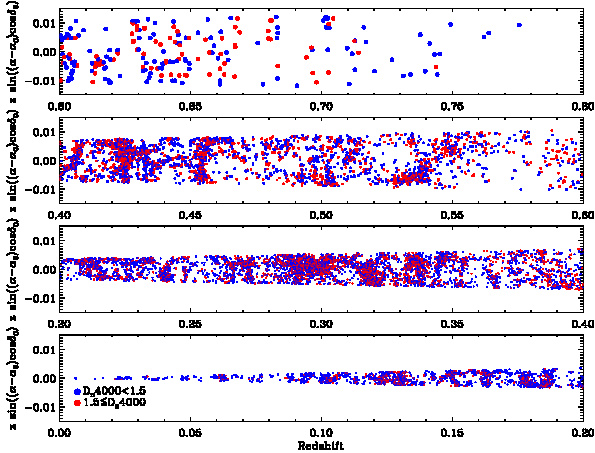}
\includegraphics[width=3.15in]{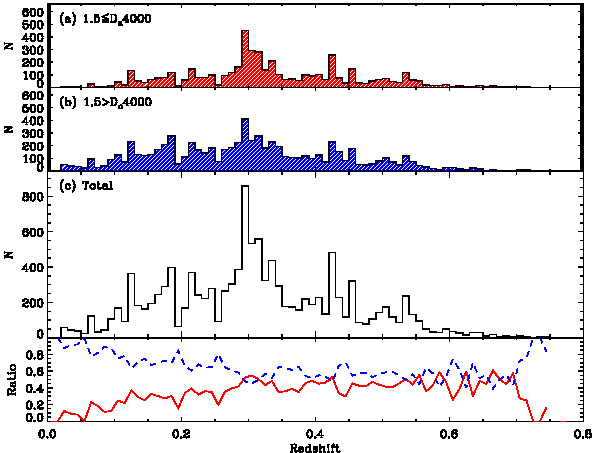}
\end{tabular}
\caption{Cone diagram (left) for the entire SHELS survey of the DLS F2 field projected in right ascension. Blue points represent galaxies with D$_n$4000 $< 1.5$ (generally star-forming); red points represent objects with D$_n$4000$\geq$1.5 (generally quiescent).
The right hand panels show redshift histograms for the population segregated by D$_n$4000
(top two panels), the total histogram (third panel) and the population fractions segregated by D$_n$4000 (bottom panel).
\label{D4000hist}}
\end{figure}

Figure \ref{D4000hist} (left-hand panel) displays the redshift survey of the F2 region projected in right ascension. The general character of the large-scale structure is obvious: there are large low density regions separated by relatively thin structures where the galaxies are. The most obvious
finger corresponding to two massive clusters of galaxies is at z = 0.2915 and z = 0.3004 corresponding to the Abell 781 complex. There are also rich clusters at redshifts z = 0.43 and z = 0.53 that correspond to 
well-defined peaks in the weak lensing map and to known extended x-ray sources (Geller et al. 2010). Fingers in redshift space corresponding to these clusters are less and less obvious at greater and greater redshift; the elongation is simply a smaller fraction of the mean redshift of the system.

A vast literature explores the relationship between galaxy properties and environment
(Blanton \& Moustakas 2009).
Galaxies may be segregated by, for example, color, an array of morphological properties,
and stellar mass. The spectral indicator D$_n$4000 (Balogh et al. 1999; Kauffmann et al. 2003) we explore briefly here is applied less often (e.g. Bundy et al. 2006; Roseboom et al. 2006; Freedman Woods et al. 2010; Moresco et al. 2010; Moresco et al. 2013), but it has the advantage that it requires no K-correction and it is insensitive to reddening.

In Figure \ref{D4000hist} we color-code the objects according to the value of  D$_n$4000 measured from the Hectospec spectrum. D$_n$4000 (Balogh et al 1999) is a measure of the amplitude of  the characteristic 4000 \AA\ break in a galaxy spectrum. The value of D$_n$4000 is a measure of the age of the aggregate stellar population (although somewhat affected by the metallicity). Freedman Woods et al. (2010) show that D$_n$4000 is well-correlated with the presence of emission lines indicative of continuing star formation. A division of the galaxy population  at D$_n$4000 = 1.5 is an approximate division between quiescent (large D$_n$4000) and star-forming objects.   

\begin{figure}
\centering
\begin{tabular}{cc}
\includegraphics[width=3.15in]{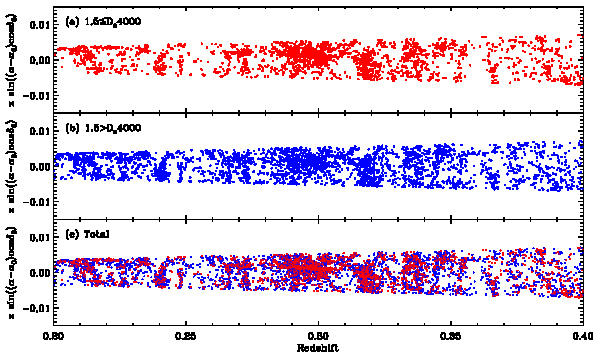}
\includegraphics[width=3.15in]{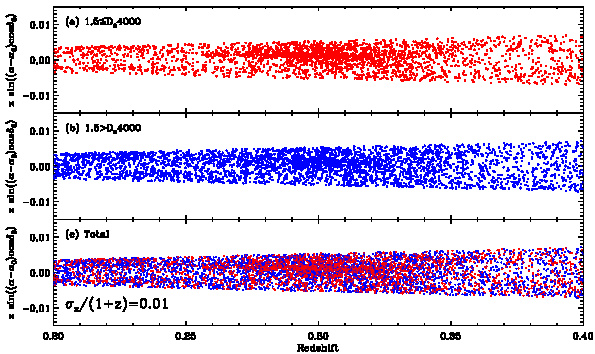}
\end{tabular}
\caption{Cone diagram for the redshift range 0.2-0.4 visually highlighting the difference
in the distribution of the populations segregated by D$_n$4000 (left). The cone diagrams in the same range based on perfect 1\% photometric redshifts reveal virtually no structure (right). The difference in the distribution of the two populations remains, but it is much less evident.  
\label{Sconephot}}
\end{figure}

Figure \ref{Sconephot} displays the redshift survey in Figure \ref{D4000hist} for the interval 0.2 $ < z < 0.4$. We show the two populations segregated by D$_n$4000 separately to highlight the difference in the spatial distribution: the older (quiescent) galaxies populate the dense central finger of the A781 cluster (and other massive systems). The contrast of the structure is greater for galaxies with larger D$_n$4000; i.e. they are more strongly clustered. Galaxies with smaller D$_n$4000 are relatively abundant in the low density voids. As expected, these effects parallel the segregation observed as a function of color and/or morphology.

The redshift histograms in Figure \ref{D4000hist} (right-hand panel) underscore these points. The lower panel shows that
galaxies containing a younger stellar population dominate the sample at low redshift. This behavior, a property of magnitude limited surveys, results from the sampling of galaxies of low intrinsic luminosity and stellar mass at low redshift (see Geller et al. 2014b). These objects tend to contain a younger stellar population than more luminous, massive systems.

A redshift survey obviously enables detailed analysis of the properties of galaxies as a function of environment and redshift. Many workers use photometric redshifts as a basis (or partial basis) to such studies in this redshift range. Figure \ref{Sconephot} (right-hand panel) shows the portion of the redshift survey in Figure \ref{Sconephot} (left-hand panel) with perfect 1\% photometric redshifts. In other words, we take the measured redshift and draw a corresponding photometric redshift from a Gaussian with
a dispersion, $\Delta{z}/(1 + z)$ = 0.01. The A781 complex remains visible, but all of the other exquisite structure in the redshift survey essentially disappears. Some evidence that quiescent (large D$_n$4000) galaxies are more clustered remains, but the signal is vastly reduced.

A complete redshift survey like SHELS remains observationally expensive. New wide-field spectrographs promise very large complete surveys perhaps beginning within the next decade. Meanwhile 
color-selected redshift surveys provide a shortcut to exploring the universe in this redshift range. 

BOSS (Baryon Oscillation Spectroscopic Survey) is a prime example of an ambitious color-selected redshift survey (Eisenstein et al. 2011; Dawson et al. 2013).  The survey maps the distribution of luminous red galaxies
with the goal of detecting the scale imprinted by the baryon acoustic oscillation in the early universe. The survey includes redshifts for some 1.5 million red galaxies covering a quarter of the sky, an average surface number density of some 150 galaxies per square degree. The survey has yielded an impressive detection of the baryon oscillation scale (e.g. Kazin et al. 2010; Percival et al. 2010; Padmanabhan 2012).

Impressive as the BOSS survey is, the sampling of the galaxy distribution to the typical depth of the survey is sparse; the number density on the sky is less than 10\% of the number of objects brighter than the typical limiting apparent magnitude regardless of color.
Thus, in contrast with a survey like SHELS, the applications of the BOSS survey are limited to large-scale features of the galaxy distribution and to investigations of the properties of the individual galaxies.

In contrast, the applications of SHELS are limited by its small areal coverage. For example, although clusters of galaxies within the survey volume are well-sampled, there are few of them. Our more extensive, but color-selected survey, HectoMAP, fills a niche between the sparse BOSS survey and the very dense SHELS survey.

\section{HectoMAP: A Large Area Dense Redshift and Weak Lensing Survey}

HectoMAP, a panoramic redshift survey, covers the region $200^\circ < \alpha< 250^\circ$ and $42.5^\circ < \delta < 44^\circ$ with a median redshift z = 0.38. The ∼ 80,000 galaxy survey to a limiting $r = 21.3$ will provide a
dense sampling of clusters and the relationship between clusters and large-scale structure
in a volume of $10^8$ Mpc$^3$ (∼ 1/7 of the volume covered by the SDSS main spectroscopic
sample). The ∼ 70, 000 redshifts in hand so far trace extensive Great Walls and voids
throughout the survey redshift range. 

HectoMAP is a substantial international project. The HectoMAP region will be covered by the HyperSuprime-Cam (HSC) Survey on the Subaru telescope. Among other goals, we expect to carry out a weak lensing study based on grizY deep imaging of the region. 
 
The very large volume Horizon Run  n-body simulations provide a basis for testing HectoMAP against
the models (Kim et al. 2011; Park et al. 2012). These dark matter only simulations cover 
the largest volumes to date. They support the extraction of many independent mock surveys derived from true light cones and thus they provide a  robust testbed for comparison with  the data.

Goals of HectoMAP include: (1) comparison of the projected mass density predicted from the foreground redshift survey with the Subaru HSC weak lensing map, (2) a direct measure of the mass accretion rate of clusters of galaxies, (3) exploration of the largest structures including the detailed relationship between massive clusters and the surrounging large-scale structure, and (4) the evolution of voids.  

\subsection {Galaxy Selection for HectoMAP}

HectoMAP is a red-selected survey. The color selection of the survey is determined by a combination of the scientific goals with the limitations imposed by very finite telescope time. In addition, to maximize the time available the HectoMAP strip is at high declination, essentially always 30$^\circ$ away from the moon. Thus we have been able to use gray and even 
bright time to observe some of the galaxies.

\begin{figure}
\centering
\begin{tabular}{cc}
\includegraphics[width=3.15in]{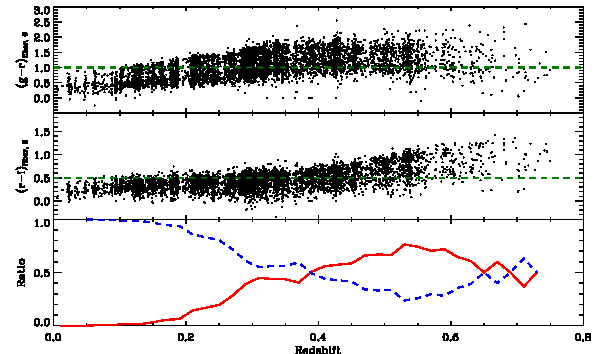}
\includegraphics[width=3.15in]{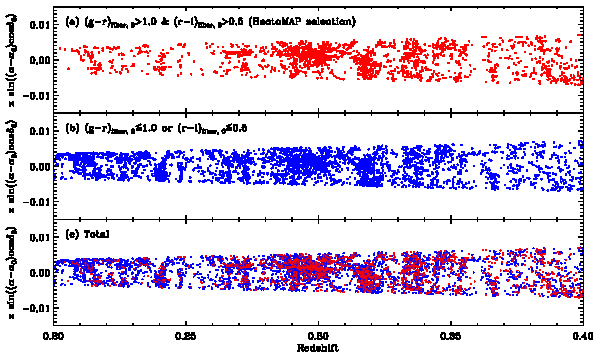}
\end{tabular}
\caption{(Left) Color-selection for HectoMAP. The upper panel shows the $g-r$ selection applied to SHELS as a function of redshift, the second panel shows the $r-i$ selection, and the third panel shows the fraction included (red) and excluded (blue). (Right) Color selected cone diagram for HectoMAP. The upper panel shows the HectoMAP
objects; the middle panel shows the bluer galaxies not observed for the main HectoMAP survey, and the lower panel shows the complete survey. The selection against galaxies with z $\lesssim 0.25$ is obvious. 
}
\label{hectoselect}
\end{figure}

Figure \ref{hectoselect} (left-hand panel)  shows the color selection adopted for the HectoMAP survey applied to the complete SHELS dataset. HectoMAP includes objects with $g-r > 1.0$ and $r-i > 0.5$. The upper panels of the figure show the color selection. The third panel shows the fraction of SHELS objects included in HectoMAP (red) and the fraction excluded (blue dashed curve).

We intentionally use $r-i$ to select against objects with $z \lesssim 0.2$. Figure \ref{hectoselect} shows the selection for a limiting R = 20.6, about 0.3 magnitudes brighter than the limit of HectoMAP. The resulting number density of objects to a limiting R = 20.6 is $\sim 1500$ per square degree.

Figure \ref{hectoselect} (right-hand panel) demonstrates the qualitative impact of the HectoMAP selection on the SHELS cone diagram in the redshift range 0.2$ <z < 0.4$.   As expected the red selection highlights clusters and enhances the apparent contrast of the large-scale structure. The very sparse sampling at the low redshift end of the range is obvious.

HectoMAP will be the first extensive dynamical analysis of clusters derived
from a redshift survey at 0.25 $\lesssim z \lesssim $0.7. At about 1/10 of the density of HectoMAP, the BOSS surveys are too sparse to address the physics of clusters of galaxies in the redshift range we propose to explore. 

\begin{figure}
\centerline{\includegraphics[width=5.0in]{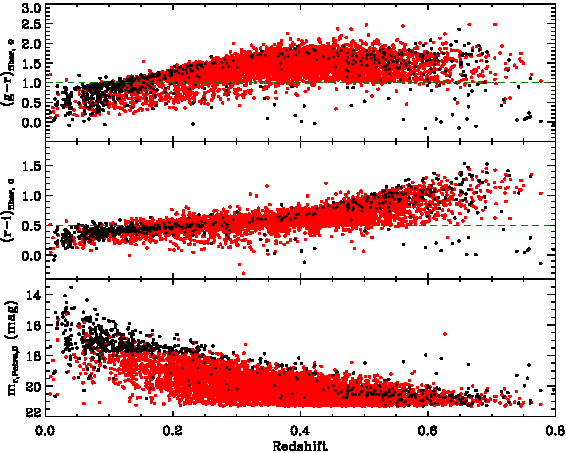}}
\vskip 5ex
\caption{BOSS selection compared with HectoMAP selection. The upper panel shows the distribution of SDSS spectroscopic objects (black) and HectoMAP objects (red). At low redshfit (z $\lesssim 0.15$) the SDSS main sample dominates. At redshifts bewteen 0.2 and 0.5 the BOSS LRG sample has a much narrower color selection than HectoMAP. At redshifts $\gtrsim 0.5$ the SDSS quasar surveys contribute bluer objects. The panels show the relative selection in $g-r$ (top), $r-i$ (middle) and $r$ (bottom). Dashed lines in each panel show relevant HectoMAP selection limits.
\label{BOSS}}
\end{figure}

Figure \ref{BOSS} shows the typical color and apparent magnitude distributions of 
spectroscopic objects in the SDSS catalog (including the BOSS surveys) as black points;
red points show the distribution of HectoMAP objects. At low redshift the main SDSS survey
with  a limiting r = 17.77 dominates the distribution of black points and the color distribution spans essentially the
full observed range. At 0.2 $ \lesssim z \lesssim 0.5$ the BOSS LRG sample with a color selection much narrower than the one we adopt for HectoMAP,  dominates the SDSS objects. At $z \gtrsim 0.5$ the color distribution for the SDSS objects broadens reflecting the inclusion of quasars.  

\subsection {A Dense Redshift Survey and a Weak Lensing Map}

Redshift surveys and weak lensing maps are two of the most powerful tools of modern cosmology. Joint applications of these tools are just developing in part because dense redshift surveys to a depth that matches the sensitivity of the weak lensing maps remain largely a dream for the future.

Geller et al. (2005) first cross-correlated projected mass maps derived from a redshift survey with a weak lensing map of one of the DLS fields. They constructed mass maps of
slices through the redshift survey by identifying systems of galaxies and then using the
square of the observed line-of-sight velocity dispersion as a proxy for the mass in a pixelized sampling of  slices in redshift space. Cross-correlation of these mass maps with the DLS weak lensing map demonstrates that the weak lensing map images the foreground large-scale structure marked by groups (and clusters)  of galaxies.

Recently van Waerbeke et al. (2013) used simulations and the recent CFHTLenS maps to explore some of the rich promise of combining large area weak lensing maps of the projected mass distribution with the foreground distribution of matter  as traced by galaxies. These investigations extend the more limited observational investigations by Geller et al. (2005, 2010). van Waerbeke et al. (2013) show that the lensing convergence maps have the potential to reveal voids in the projected matter distribution. This application was previewed observationally by Miyazaki et al. (2002) and by Geller et al. (2005).

The use of photometric redshifts for  characterizing the foreground galaxy distribution is a potentially important limitation of the techniques van Waerbeke at al (2012) investigate.
As Figure \ref{Sconephot} shows, photometric redshifts erase the details of large-scale structure that can contribute to the lensing signal. Geller et al. (2013) discusses the particularly striking example of two clusters (Abell 750 and MS0906+11) superimposed along the line-of-sight and contributing comparably to the weak lensing mass. Photometric redshifts would not resolve the two clusters. A spectroscopic survey like HectoMAP is an important testbed for methods of alleviating these limitations and for beginning to understand the statistics superimposed structures along the line-of-sight from an empirical, observational point of view.

In addition to the full exploration of the characteristics of the entire weak lensing map, individual peaks in the weak lensing map identified with massive clusters of galaxies  probe the distribution of cluster masses. So far, the samples of peaks in existing maps present some puzzles. Even peaks at high significance (i.e. with a signal-to-noise greater than 3) are often spurious (Geller et al. 2010; Kurtz et al. 2012; Utsumi et al. 2014; Shan et al. 2012, 2014). The combination of a large area dense redshift survey with a weak lensing map should be a route to a better understanding of the nature of the weak lensing map peaks.

Based on the SHELS survey, we estimate that HectoMAP should include
 ∼ 200 clusters with restframe line-of-sight velocity dispersion $\gtrsim$ 600 km s$^{-1}$.
Comparison of this
sample from HectoMAP with the Subaru weak lensing maps will obviously provide a powerful measure
of the correspondence between the amplitude of the weak lensing signal and the central
velocity dispersion of the clusters. HectoMAP will complement larger weak lensing cluster samples based largely on photometric redshifts and scaling relations applied to photometric properties of candidate cluster members (e.g. Shan et al. 2012; 2014). As Geller et al. (2013)  and Hwang et al. (2014) demonstrate, even the best available photometric redshifts cannot resolve structures along the line-of-sight that contribute to a weak lensing signal. 

Comparison of the mass distribution marked by the HectoMAP redshift survey with the Subaru HSC weak lensing map is a unique opportunity to test two large-scale, independent measures of the
mass distribution in the universe against one another. As techniques for constructing the maps and for using the redshift survey improve, so will our understanding of the way galaxies mark the matter distribution in the universe.

\subsection{Comparing HectoMAP with the Horizon Run 2 Simulations}
\label{simulation}

HectoMAP is designed to sample a large range of physical scales and to intersect many ``voids'' at redshifts $\gtrsim 0.25$. Testing the data against numerical simulations is
always challenging. In our first approach to a test we match the data to the
Horizon Run 2 N-body simulation (Kim et al. 2011). This simulation follows the evolution of the distribution of 6000$^3$ dark matter particles in a huge comoving volume of 10$^{3}$ Gpc$^3$.

\begin{figure}
\centering
\begin{tabular}{cc}
\includegraphics[width=3.15in]{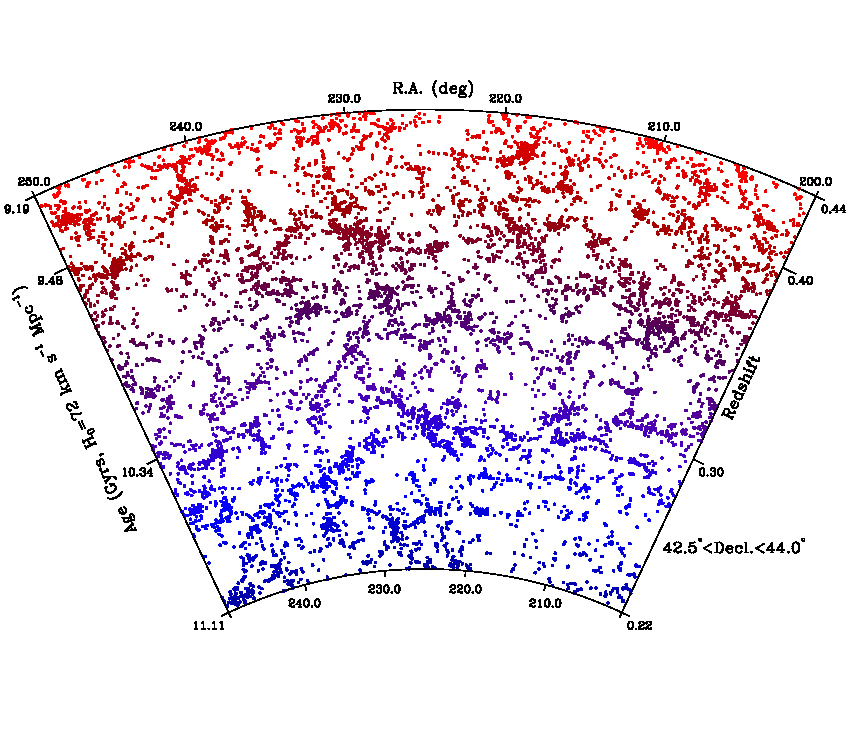}
\includegraphics[width=3.15in]{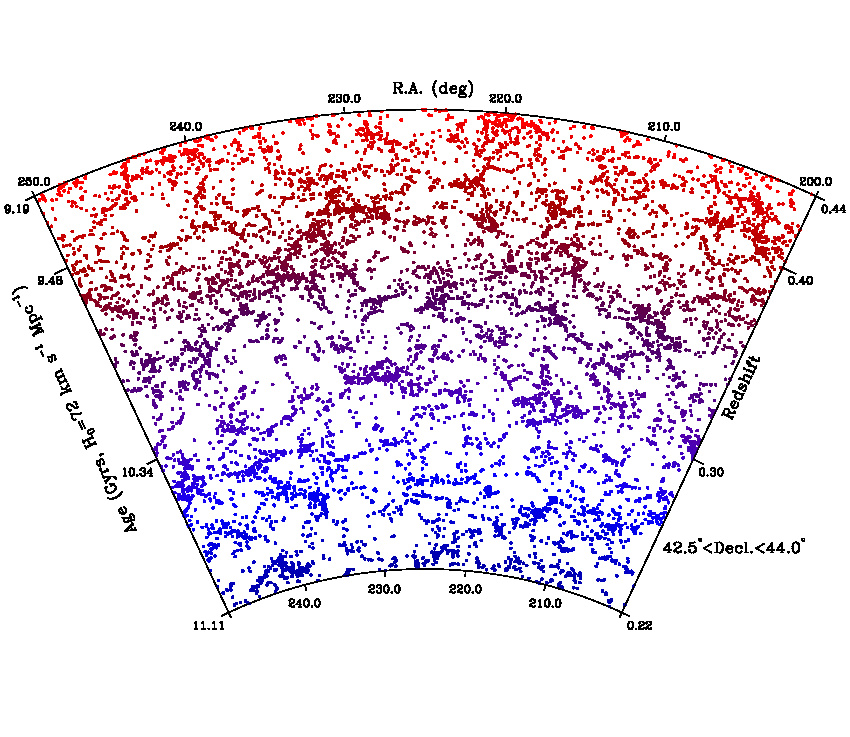}
\end{tabular}
\caption{Left: A volume-limited sample of HectoMAP data for $r<20.5$.
Note the obvious fingers corresponding to clusters of galaxies.
A large structure appears to cross the survey at $z\sim0.3$.
(Right) A true lightcone from the Horizon Run 2 simulation 
  (Kim et al. 2011) with the selection function
  for the data applied.
Note the apparently less empty voids, and relatively poor sampling of clusters of galaxies.
}
\label{hectomap}
\end{figure}

To compare the data with the simulations we use the HectoMAP data for r$\leq$ 20.5 where the survey is now complete. We extract a volume-limited sample covering the redshift range 0.22$ < z < 0.44$
(Figure \ref{hectomap}, left).
The sample has a constant galaxy number density and contains galaxies with a K-corrected (to $ z = 0.4$) absolute magnitudes M$_r \lesssim -20$. We test this map against 280 HectoMAP-like mock surveys based on true light cones in the simulation. The minimum halo mass we sample is 4.3$\times$10$^{12}$ M$_\odot$. The typical galaxy (halo) separation for the least massive objects is $\sim 12.5$ Mpc. The right panel of Figure \ref{hectomap} shows a typical simulated sample. The visual similarity is impressive although there are some subtle differences including (1) the appearance of ``fingers'' in the data and (2) the apparently emptier voids in the simulation.

We have applied a friends-of-friends algorithm  to the data and the simulated samples. Impressively, in concert with recent results of the GAMA survey
(Alpaslan et al. 2014), our preliminary results show that large dense structures in the data
and in the Horizon Run 2 simulations are similar. Interestingly, the size distributions of voids are only marginally consistent at this depth; there is an excess at the $\sim 2 \sigma$ level voids with diameters
$\gtrsim$ 50 Mpc. 

Completion of the survey to $r = 21.3$ (expected in the spring of 2015) will more than double the number density of galaxies in the volume limited sample shown in Figure \ref{hectomap}. With this denser
sample and with the much denser Horizon  Run 4 Simulations, we will be able to explore any
systematic issues that might account apparent differences in the distribution of void sizes. 

The survey to $r = 21.3$ will enable construction of more extensive samples analogous to those in Figure \ref{hectomap}. These maps can, in principle, sample larger structures; they will thus also test
the conjecture (Park et al. 2012) that the largest dense structures should
extend for a maximum of 430 Mpc. The complete HectoMAP will access more than three times
this scale.

There are few direct observational limits on the evolution of voids. Based on DEEP, Conroy et
al. (2005) do show that voids grow over cosmic time as expected.
Micheletti et al. (2014) examine voids in the VIPERS survey over the redshift range
$0.5 < z < 1.2$ and construct a void catalog. However, selection effects in VIPERS prevent measurement of the void size distribution function. HectoMAP is a shallower survey with a cleaner, denser selection. In the range where both VIPERS and HectoMAP surveys are useful for
exploring the nature of voids, the HectoMAP survey covers a significantly broader range of comoving physical scales: at $z = 0.55$, for example, the W1 field of VIPERS has transverse comoving dimensions of 308 $\times$ 47 Mpc whereas the HectoMAP dimensions are approximately 1.3 Gpc $\times$ 55 Mpc. Thus we expect the completed HectoMAP will provide reasonably robust estimates of the void size distribution function and its evolution to a redshift 
$z \simeq 0.7 - 0.8$.

The combination of a foreground dense redshift survey with a weak lensing map holds promise for probing the distribution of dark matter in voids. In a portion of CFHTLenS, van Waerbeke et al. (2013) use photometric redshifts to construct  a mass map that they compare with the weak lensing projected mass map. They show that troughs in the two maps correspond impressively well, underscoring the possibility of investigating voids with a combination of weak lensing
and foreground galaxy surveys. Photometric redshifts as opposed to a spectroscopic survey limit the van Waerbeke approach in ways that are hard to evaluate without an
observational test based on a dense spectroscopic survey. In the next couple of years, HectoMAP will enable a comparison of measured proxies for the matter distribution based on spectroscopy and on weak lensing. Like our first steps toward mapping the distribution of galaxies in the nearby universe, HectoMAP is part of the process of understanding the nature and evolution of structure in the universe.

\section {Conclusion}

At the beginning of my career I was inspired by scientists like Martin Schwarzschild and Jim Peebles who laid the foundations for their fields. At this point in my career I feel fortunate to have played a central role in uncovering and defining what we now call the cosmic web.

During my career, advances in our ability to observe and to model the universe and its evolution have been awe-inspiring. When we mapped our first slice of the universe is 1986, I could not have imagined that today I would be able to carry out a survey like HectoMAP.  In spite of this richness, fundamental questions remain unanswered. We have learned a lot about the distribution of dark matter, but we still have no idea what it is. The nature of the mysterious dark energy is an equally deep unsolved problem. It is my fond hope that some of the extraordinarily talented people I have trained will solve these profound problems.

I am, however, concerned that the social aspects of our profession have not improved in the
way that our knowledge of the universe has. There is no doubt that we will make bigger better maps of the universe. However, the promise that the social aspects of our profession  will improve is not at all guaranteed. Access for women and for members of other traditionally underrepresented groups in the physical sciences has improved in some respects,
but there are new problems. For example, the continual lengthening time to the PhD and the lengthening post doctoral period hobble many young scientists' lives. 

In my opinion, it is incumbent on the leaders of the field to think deeply about the broad range of issues affecting young scientists. The overproduction of astronomy PhD's is one of several serious issues. The fraction of people who get tenure-track positions is now below 25\%. In my opinion, that fraction is irresponsibly low. If we want to preserve the attractiveness and the exciting intellectual atmosphere of our field, we must consider the tacit promises we make and the disappointment we cause.

Throughout my career I have been surrounded by able younger scientists. Perhaps more than any other aspect of my career, I have enjoyed seeing many of them thrive. I have tried to give them courage to question cherished beliefs whatever they do in life. I have encouraged them to have balance in their lives and to treat their own proteges with a generosity of spirit that encourages creativity. 

There is an artistry in nature that should inform the
way we structure our profession. Inattention to the romance and the beauty that underlies what we do every day is, in my opinion, a loss of the central idea that science is a special way of seeing
the universe around us.  

The new work on HectoMAP previewed here will be published in full elsewhere. In advance. I thank Daniel Fabricant, Juhan Kim, Michael Kurtz, Changbom Park, Satoshi Miyazaki, and Yousuke Utsumi for the pleasure of working with them and for their contributions to the project. Perry Berlind, Mike Calkins,
and Susan Tokarz operated the Hectospec and reduced the data. I thank Scott Kenyon for his unfailing support and wise counsel and Andreas Burkert for an insightful reading of this manuscript. The Smithsonian Institution and the Korean Institute for Advanced Study have generously supported the research discussed here.

\clearpage

\clearpage

\clearpage

\clearpage

\end{document}